\documentclass[a4paper,11pt]{article}
\pdfoutput=1 

\usepackage{jinstpub} 
\usepackage{subfigure}
\usepackage{lineno}

\title{\boldmath The BIS78 Resistive Plate Chambers upgrade of the ATLAS Muon Spectrometer for the LHC Run-3}

\author[a]{L. Massa,\footnote{Corresponding author.} }

\affiliation[a]{INFN, Sezione di Bologna, Italy}

\emailAdd{lorenzo.massa@bo.infn.it}

\abstract{Resistive Plate Chambers (RPCs) are used in the ATLAS experiment for the muon trigger and two coordinate measurements in the central region.

In preparation for the coming years of LHC running at higher luminosity, besides the New Small Wheel project which is expected to complement the ATLAS muon spectrometer in the end-cap regions, a smaller size project, known as BIS78 (from Barrel Inner Small sectors), is being developed with a foreseen installation during the LHC Long Shutdown 2 (2019-2020). The BIS78 project will reinforce the fake muon rejection and the selectivity of the muon trigger in the transition region between the ATLAS barrel and end-caps by adding 32 RPC triplets on the edges of the Inner Barrel even sectors (BIS7 and BIS8) as this region is characterized by high rate due to secondary charged tracks generated by beam halo protons and a lack of detector instrumentation. Each triplet will be a station composed by three independent RPC detectors.

Due to the narrow available space, the project foresees to replace the existing Monitored Drift Tubes (MDTs), used for the precise position measurement in this area, with muon stations formed by integrated smaller diameter tubes (sMDT) and a new generation of RPCs, capable of withstanding the higher rates and provide a robust standalone muon confirmation. 

These new RPCs are based on novel design of the gas volume with a thinner gas gap, thinner resistive electrodes, a lower operating voltage and new high gain front-end electronics with respect to the legacy ATLAS RPCs. 

Besides the use in Run-3 and onwards, this project is also of particular relevance as a pilot test for the High Luminosity LHC upgrade when an additional full layer of new RPC triplets is expected to complement the full barrel region in the innermost plane. 

The core of the project is presented, together with a description of the production and the test results. Details on the detector infrastructure and services along with a roadmap towards the final installation and commissioning during the Long Shutdown 2 (2019-2020) are also discussed.
}

\keywords{Gaseous detectors, resistive plate chambers, trigger detectors}

\arxivnumber{2004.12693} 

\collaboration[c]{on behalf of ATLAS Muon collaboration}

\proceeding{XV$^{\text{th}}$ Workshop on Resistive Plate Chambers and Related Detectors  RPC2020\\
  10-14 February. 2020\\
  Rome, Italy}

\begin{document}
\maketitle
\flushbottom

\section{The ATLAS muon spectrometer}
\label{sec:intro}
ATLAS \cite{ATLAS} is a multipurpose experiment at the Large Hadron Collider (LHC), which covers almost completely the whole solid angle, using a large number of sub-detectors. 
In particular, the muon spectrometer \cite{spectrometer}, situated in the outermost region of ATLAS, is crucial in all searches for new physics. 
It is a huge apparatus (46 meters long and 25 meters high) made of almost 4000 detectors, where the muon trajectory is bent by a magnetic system composed by the inner-detector solenoid field of 2 T and a toroidal field of $\sim 0.5$ T ($\sim 1$ T) in the barrel (end-caps).

The tracking of the particles is performed with two kinds of precision detectors: Monitored Drift Tubes (MDTs) and Cathode Strip Chambers (CSCs).
On the other hand, the passage of the particle is triggered using other two types of faster detectors: the barrel region (|$\eta$|<1.05) is covered by Resistive Plate Chambers (RPCs) while the end-cap region  (|$\eta$|>1) is covered by Thin Gap Chambers~(TGCs).\footnote{The pseudorapidity $\eta$ is defined as $\eta=-\ln\left(\tan\frac{\theta}{2}\right)$, where $\theta$ is the polar angle.}

\subsection{The ATLAS RPC system}
The RPCs provide the muon first level trigger signal in the barrel region of the spectrometer, covering a total surface of 4000 m$^2$, with more than 3600 gas volumes.
The present system is organized into three concentric layers of RPC doublets: two layers are in the Barrel Middle (BM) region, one layer is in the Barrel Outer (BO) region, while no layer of RPCs is installed in the Barrel Inner (BI) region (its installation is foreseen for the ATLAS Phase-II upgrade \cite{phase2tdr}).

The trigger algorithm is based on RPC hit coincidence: the low-$p_\textrm{T}$ trigger ($p_\textrm{T}$< 10 GeV) asks for coincidence between the two layers in the BM region (with a signal in at least 3 gas volumes out of 4), while the high-$p_\textrm{T}$ trigger ($p_\textrm{T}$>10 GeV) requires an additional confirmation in one BO station.

\section{ATLAS Phase I upgrades}
\label{sec:project}
\subsection{High rate in forward region}
In the |$\eta$|>1 region of the spectometer a high fake trigger rate is observed, due mainly to low-$p_\textrm{T}$ protons emanating from the end-caps toroids and shieldings. In particular, the amount of fake trigger signals is expected to rise in ATLAS in Run-4, when the LHC luminosity will be up to $L= 7.4\cdot10^{34}$ cm$^{-2}$ s$^{-1}$, more than 3 times larger than the current luminosity. In order to cope with the new conditions, the muon spectrometer is being upgraded  \cite{tdaqtdr}.

To reduce the trigger rate in the forward region (|$\eta$|>1.3) the New Small Wheel (NSW) will be installed for ATLAS Run-3 \cite{nswtdr}.

However in the region between the barrel and end-caps of the spectrometer (1<|$\eta$|<1.3) the fake trigger rate will not be reduced by the NSW and ATLAS plans to improve the trigger capability by adding new detectors.

The large sectors of the spectrometer's end-caps are already equipped with TGCs, which can be used to reduce half of the remaining fake rate. In the other sectors, where the toroid is placed, there is no room for detecors in the end-cap, and the rate reduction will be obtained by adding a new layer of RPCs in the barrel.
\subsection{The BIS78 upgrade project}
\begin{figure}[!h]
\centering
\subfigure{\includegraphics[width=7.7cm]{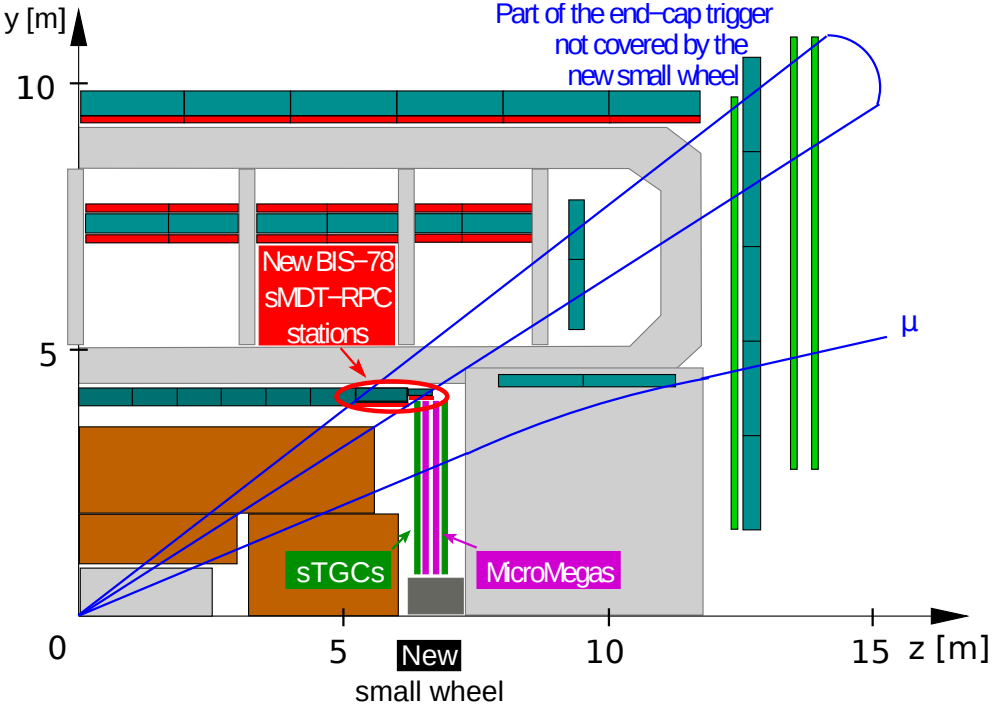}}
\subfigure{\includegraphics[width=7.3cm]{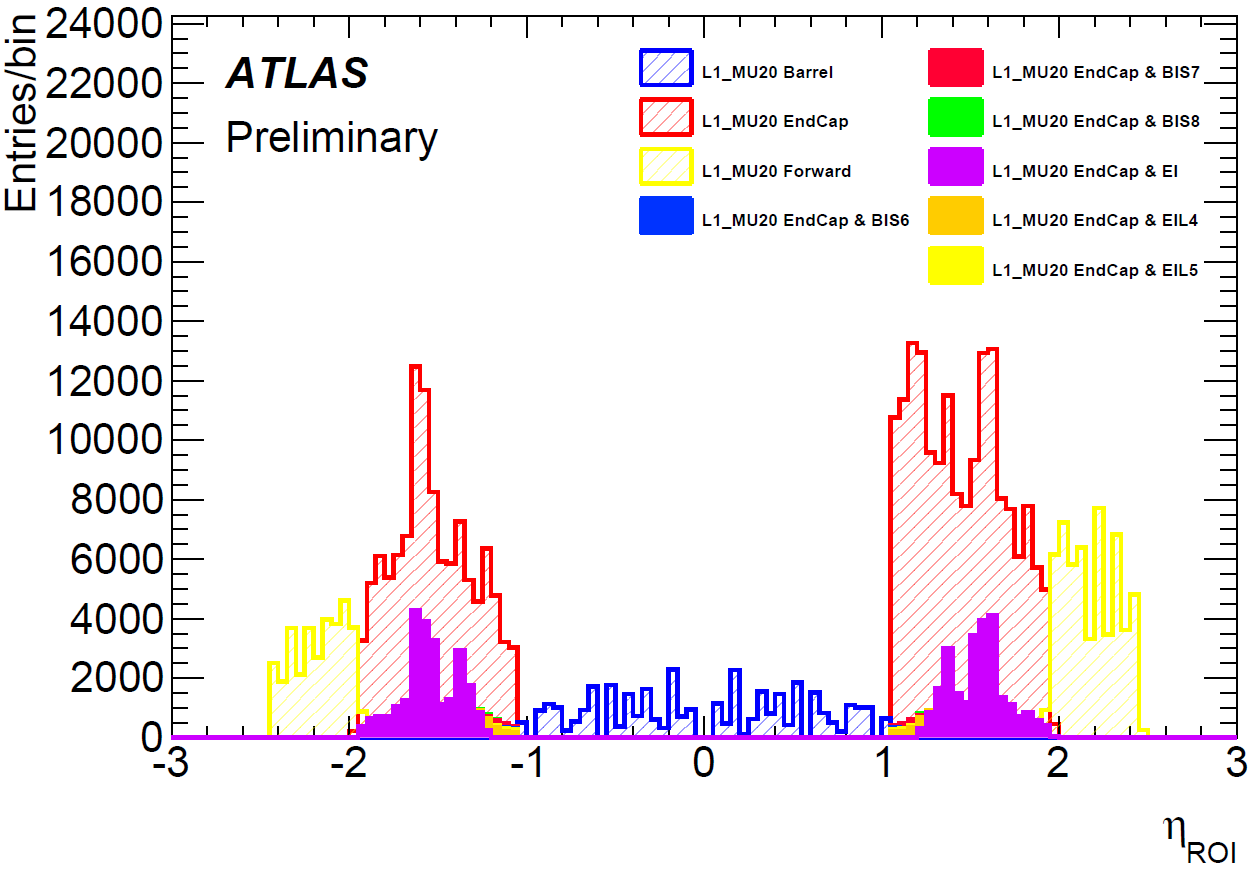}}
\caption{\small{On the left: partial side view of the ATLAS muon spectrometer, in the transition region between the barrel and end-caps. The position of the BIS78 stations is circled in red. On the right: rate reduction obtained adding BIS chambers, observed in the $\eta$ distribution of the ATLAS triggered events ($p_\textrm{T}$>20 GeV). The hatched distribution is obtained with the current end-cap trigger, while the fully colored one is obtained requiring the passage through the end-cap TGCs or BIS chambers \cite{ratereduction}. Copyright 2020 CERN for the benefit of the ATLAS Collaboration. CC-BY-4.0 license.}}\label{fig:geometrical_coverage}
\end{figure}

This so-called BIS78 project is considered as a pilot for the Phase-II BI RPC upgrade \cite{phase2tdr}, to test the performances of the new generation of RPCs.
Indeed, the existing 32 BIS7 and BIS8 MDT will be replaced by 16 new muon stations made of one smaller MDTs (sMDT) BIS7/8 chamber and two RPC triplets (BIS7 and BIS8). 
The first 8 stations, corresponding to one side of ATLAS, will be installed already in 2020, while the rest will be installed during the third Long Shutdown of LHC. The position where the BIS78 stations will be installed and their effect on the fake rate reduction can be seen in Figure~\ref{fig:geometrical_coverage}.

\section{New generation of RPCs}
\label{sec:newgen}
An intense R\&D activity has been carried out to develop a new generation of RPCs with the performance and geometrical constraints required for the BIS78 triplets.
These new RPCs will have thinner gas gaps (1 mm instead of 2 mm) and thinner electrodes (1.3 mm instead of 1.8 mm), leading to several improvements in terms of weight and size
of the detector, time resolution, signal collection efficiency and charge distribution. Moreover, the new RPCs will work at almost half of the applied high voltage with respect to present RPCs, maintaining a similar electric field.

The front-end electronics of these new generation of RPCs integrates a new amplifier in silicon \cite{ampli} and a new discriminator in SiGe BiCMOS technology \cite{performances}, leading to higher rate capability, radiation hardness, better space time resolution and inexpensive high performance with reduced power consumption. The main features of the new amplifier and discriminator Application Specific Integrated Circuits (ASICs) are summarized in Table~\ref{tab:amplifier_discriminator}.

It has been proved that this generation of RPC reached an outstanding performance, in particular a space resolution of 1 mm and a time resolution of 0.4 ns \cite{performances}.

\begin{table}[!h]
\begin{center}
\caption{\small{Main features of the new fast charge amplifier ASIC implemented in Si BJT technology (left) and of the new discriminator ASIC implemented in SiGe BiCMOS technology (right).}}
\vspace{3mm}
\resizebox {0.4\textwidth}{!}{
  \begin{tabular}{|cc|}
  \hline
  \multicolumn{2}{|c|}{Amplifier (Si)}\\
\hline
\hline
Gain &0.2-0.4 mV/fC\\
Power consumption& 3-5 V, 1-2 mA\\
Band width &100 MHz\\
\hline
  \end{tabular}}
 \qquad
 \resizebox {0.4\textwidth}{!}{
  \begin{tabular}{|cc|}
  \hline
  \multicolumn{2}{|c|}{Discriminator (SiGe)}\\
\hline
\hline
Threshold &0.5 mV\\
Power consumption& 2-3 V, 4-5 mA\\
Band width &100 MHz\\
\hline
  \end{tabular}}
  \label{tab:amplifier_discriminator}
\end{center}
\end{table}

\section{BIS78 production at CERN}
\label{sec:productionCERN}
The raw materials used for RPCs are produced jointly in Italy (gas volumes, copper strips for readout panels, front-end boards) and Germany (front-end ASICs and mechanical frames). Everything is then shipped to CERN where the RPC triplets for the BIS78 project are finally assembled and tested through a series of quality assurance and control (QA/QC) procedures, explained in the following sections.

At CERN, the gas gaps are irradiated with a gamma ray source and tested under harsher conditions with respect to the photon flux of 600 Hz/cm$^2$ expected at HL-LHC (section \ref{sec:gasgap}).
At the same time, the copper strip panels are glued on the forex insulator to make the readout panels, and the front-end boards are tested (section \ref{sec:frontend}). 

Then, the front-end boards are solded on one side of the readout panels, while on the other side  matching resistors are used in order to avoid signal reflections inside the copper strips. 

Once their quality is checked by measuring their working parameters (like the ohmic current and the knee point for the gaps and the counting rate and the current drained by the low voltages for the boards), gas gaps and readout panels are coupled to make the RPC singlets, which are tested with cosmic rays (section \ref{sec:cosmicray}).
The singlets are then integrated into the mechanical frames to make triplets, which are then cabled and integrated with the sMDT before the final cosmic ray test.

\subsection{The gas gap QA/QC}
\label{sec:gasgap}
The gas gaps used for the BIS78 project have to face a first acceptance test at General Tecnica Engineering, the site where they are produced in Italy. 
There, an accurate visual inspection for surface defects is done on the bakelite plates used to make the gaps, together with measurements of the graphite electrode surface resistivity (500 $\Omega /\Box \pm 40\%$) and the spacer thickness tolerance of 15 $\mu$m. On some samples, the spacer gluing reliability is checked with a disruptive test applying a minimum strength of 3 kg.
After the gas gaps are assembled, the polymerization of the oil coating the resistive plate inner surface is checked through a destructive test, opening dedicated samples of small dimensions which are built in parallel. 
A visual inspection is then performed on the external surface of the gaps to check the glueing between the external PET insulator and the electrodes, verifying the absence of air bubbles.
All the gaps have then to face tests on gas tightness and high voltage insulation, and their Volt-Amperometric characteristics are measured. 

The gaps are then shipped at CERN, where they are tested with one week of gamma irradiation. When the gaps are exposed to irradiation, their current rises as a consequence of the small avalanches induced by the random interaction with the incoming photon flux. During this week, the gaps follow the conditioning process, being slowly turned on at high rate: they all reach 40 $\mu$A in steps of 10 $\mu$A, which take around half a day each.

\begin{figure}[!h]
\centering
\subfigure{\includegraphics[width=7.1cm]{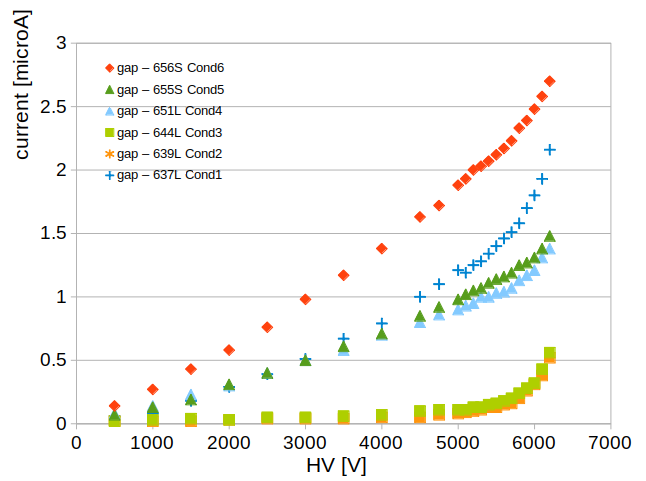}}\qquad
\subfigure{\includegraphics[width=7.1cm]{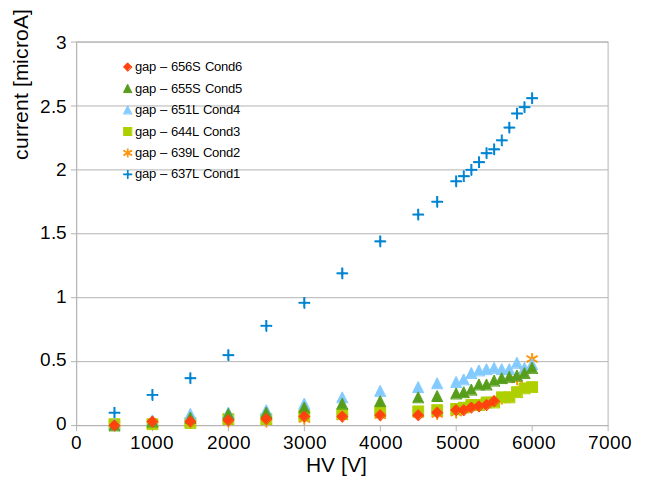}}
\caption{\small{Volt-Amperometric curves of some gas gaps used for BIS78 production, measured before (left) and after (right) one week of conditioning with gamma irradiation. 
After conditioning, both linear (ohmic) and exponential parts of the curves are drastically reduced in all the gaps except 637L, which has not been used to produce any RPC singlet. Copyright 2020 CERN for the benefit of the ATLAS Collaboration. CC-BY-4.0 license.}}\label{fig:conditioning}
\end{figure}

The positive effects of this process can be seen in Figure \ref{fig:conditioning}, where  the Volt-Amperometric curves of some gaps measured before the conditioning procedure are compared with  the curves of the same gaps after one week of conditioning. 
Both the linear (ohmic) and exponential parts of the curves are drastically reduced by the conditioning, burning all the small defects on the surface of the gas gaps. Just one gas gap has not showed this positive effect, hence it has been discarded. 
Until now, 45 gas gaps have been tested and only 3 of them were not fulfilling the QA/QC criteria and have been discarded.

\subsection{The front-end boards QA/QC}
\label{sec:frontend}
The new front-end boards are tested carefully at CERN, varying their threshold voltage parameter $V_\textrm{TH}$. For each $V_\textrm{TH}$ value, the eight input channels are perturbated with a metal spring probe connected to a signal generator and the output signals are acquired through a Time to Digital Converter (TDC).
The front-end boards fulfill the QA/QC test when all their input channels work while changing the threshold within a range of at least 200 mV (1.5 V $\leq V_\textrm{TH} \leq$ 1.7 V) and if no crosstalk is observed.
With these criteria, 650 front-end boards have been already tested, accepting 90\% of them and obtaining just enough material to assemble the RPCs that will be installed in 2020.

\subsection{Cosmic ray tests on triplets}
\label{sec:cosmicray}
All the RPC singlets and triplets assembled at CERN are tested using cosmic rays.
The cosmic muons are triggered using two layers of scintillators covering an area of $20 \times 60$~cm$^2$.
The events are recorded on a computer using a CAEN TDC with 100 ps of time resolution, every time a coincidence between the two scintillators is observed within a time window of 100 ns.
The singlets are always tested in batches of three, and during the data taking, one singlet is left at fixed high voltage, as reference.

Singlets and triplets are accepted if they have an efficiency greater than 95\% in the plateau region, less than 1 Hz/cm$^2$ of noise, less than 1\% of dead channels and a cluster size smaller than 3 strips, compatible with the readout panel strip pitch of 25 mm.

\begin{figure}[!h]
\centering
\subfigure{\includegraphics[height=5cm]{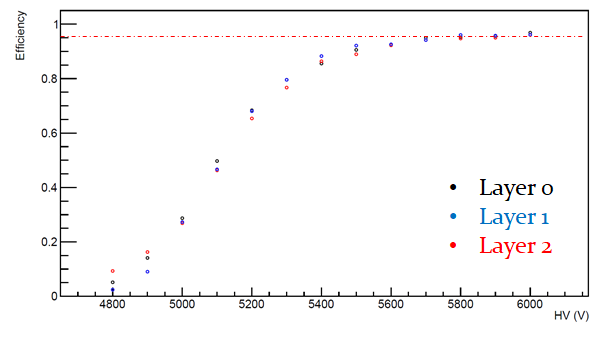}}\qquad
\subfigure{\includegraphics[height=5cm]{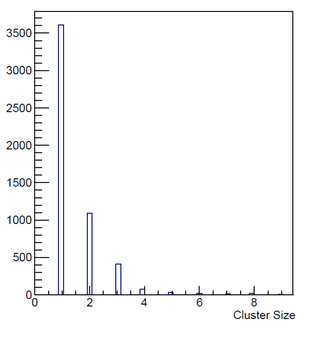}}
\caption{\small{Left: the efficiency curve measured on the BIS78 Module 0 RPC triplet. Dots of different color refer to the different singlets contained in the triplet. Each efficiency point has been calculated taking 1000 triggered events at the relative high voltage. At the applied voltage of 5.8 kV the Module 0 singlets cross the horizontal red line marking the 95\% efficiency.
Right: the cluster size measured on one RPC singlet of the BIS78 Module 0 at 5.8 kV; the observed average cluster size is 1.5 with a standard deviation of 1.1. Copyright 2020 CERN for the benefit of the ATLAS Collaboration. CC-BY-4.0 license.}}\label{fig:module0}
\end{figure}

In May 2019 the Module 0, consisting of a BIS7 RPC triplet, was integrated into the mechanical frame and tested with cosmic rays, showing an efficiency greater then 95\% at 5.8 kV of applied high voltage, an average cluster size smaller than 3 strips and no dead channels (see Figure \ref{fig:module0}). Moreover, the triplet has been integrated with the corresponding sMDT station, and no interference has been observed between the two detectors.

\begin{figure}[!h]
\centering
\includegraphics[height=5cm]{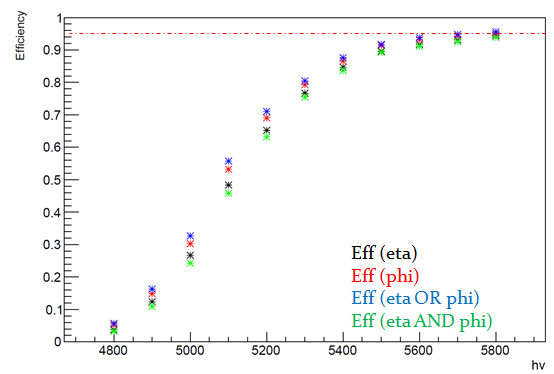}
\caption{\small{Efficiency curve measured on one BIS7 RPC singlet. The efficiency  calculated from signals from the $\eta$ front-end boards is drawn in black; the efficiency  calculated from signals from the $\phi$ front-end boards is drawn in red; the efficiency calculated making the logical OR of the signals from $\eta$ and $\phi$ front-end boards is drawn in blue, while the one calculated making the logical AND of the signals is drawn in green. Each efficiency point has been calculated taking 3000 triggered events at the relative high voltage. At the applied voltage of 5.6 kV the singlet crosses the horizontal red line marking the 95\% efficiency. The same results have been observed on all the RPC singlets produced until now. Copyright 2020 CERN for the benefit of the ATLAS Collaboration. CC-BY-4.0 license.}}\label{fig:a04}
\end{figure}

Until February 2020, three other BIS7 and BIS8 triplets have been assembled, confirming the same performance as Module 0, 
as can be seen in Figure \ref{fig:a04}, showing the results obtained with the last BIS7 triplet which has been tested. In January 2020, one complete station made of BIS7 and BIS8 RPCs has been integrated with the corresponding sMDT. 

\section{Trigger and data acquisition status}
\label{sec:trigger}
Once the BIS78 RPCs are installed in the ATLAS cavern, the signals from the front-end boards are digitized through  High Performance TDCs (HP TDCs)~\cite{hptdc} with 200 ps time resolution.
The hit data are then collected by the BIS78 Pad, an FPGA-based (Xilinx Kintex 7 family~\cite{fpga}) board performing a local trigger coincidence with a 2/3 majority logic to select a muon candidate passing through three RPC gas gaps. 
The selected events are then recorded by means of FELIX (FrontEnd Link Interface eXchange \cite{felix}), which is the data acquisiton board officially developed for the ATLAS Phase-II upgrade. 
A prototype version of the data acquisition system, complete with HP TDC, Pad and FELIX boards has been tested, reading successfully an entirely cabled BIS7 triplet and sending data to FELIX at 320 Mb/s.

\section{Conclusions}
\label{sec:conclusions}

The BIS78 project will provide 16 new integrated sMDT+RPC stations, to be installed in the transition region between the barrel and end-caps of the muon spectrometer (1.0<|$\eta$|<1.3) as part of the Phase-I ATLAS upgrade.

These stations make use of a new generation of RPCs, with a better space time resolution and rate capability with respect to the present ATLAS RPCs.

Six production triplets have been already assembled and tested, showing all an efficiency greater then 95\% in the plateau region.
At the same time, a preliminary version of the trigger and data acquisition system has been tested successfully taking data from a BIS7 triplet.

The installation in the ATLAS cavern of the first 8 BIS78 stations of the project will be performed before the start of ATLAS Run-3, while the second 8 BIS78 stations will be installed during the third Long Shutdown of LHC.

\acknowledgments

The author acknowledges the support of the AIDA-2020 project which has received
funding from the European Union's Horizon 2020 Research and Innovation programme
under Grant Agreement No. 654168.


\end{document}